\documentclass[a4paper,12pt,reqno,superscriptaddress,nofootinbib]{revtex4}
\usepackage[centertags]{amsmath}
\usepackage{amsfonts}
\usepackage{amssymb}
\usepackage{amsthm}
\usepackage{newlfont}
\usepackage{stmaryrd}
\usepackage{mathrsfs}
\usepackage{mathtools}
\usepackage{euscript}
\usepackage{graphicx}
\usepackage{enumerate}
\usepackage{todonotes}

\usepackage{color}

\usepackage{tikz}
\usepackage{pgf}
\usetikzlibrary{positioning,fit,calc}
\usetikzlibrary{arrows,automata}
\usepackage{wrapfig}
\usepackage{subfigure}
\usepackage{amscd}
\usepackage{hyperref}

\usepackage{changes}

\theoremstyle{plain}

\theoremstyle{definition}

\theoremstyle{remark}

\let\de=\delta 
\let\ve=\varepsilon  \let\ga=\gamma 
  \let\om=\omega 
\let\si=\sigma

\let\De=\Delta

\newcommand{\be}{\begin{equation}}
\newcommand{\en}{\end{equation}}

\def\om{\omega}

\let\si=\sigma

\let\De=\Delta   
 
\let\om=\omega

\newcommand{\caP}{{\mathcal P}}
\newcommand{\caQ}{{\mathcal Q}}

\newcommand{\caW}{{\mathcal W}}

\newcommand{\opunit}{\text{1}\kern-0.22em\text{l}}

\newcommand{\id}{\textrm{d}}

\DeclareMathAlphabet{\mathpzc}{OT1}{pzc}{m}{it}

\let\oldsqrt\sqrt
\def\sqrt{\mathpalette\DHLhksqrt}
\def\DHLhksqrt#1#2{%
	\setbox0=\hbox{$#1\oldsqrt{#2\,}$}\dimen0=\ht0
	\advance\dimen0-0.2\ht0
	\setbox2=\hbox{\vrule height\ht0 depth -\dimen0}%
	{\box0\lower0.4pt\box2}}

\let\de=\delta 
\let\ve=\varepsilon  \let\ga=\gamma 
  \let\om=\omega 
\let\si=\sigma   
 
\let\be=\beta

\let\De=\Delta

\DeclareMathAlphabet{\mathpzc}{OT1}{pzc}{m}{it}

\def\bea{\begin{eqnarray}}
\def\eea{\end{eqnarray}}
\def\ba{\begin{array}}
	\def\ea{\end{array}}

\begin{document}

\title{Nonequilibrium calorimetry}

\author{Christian Maes}
\affiliation{Instituut voor Theoretische Fysica, KU Leuven, Belgium}
\author{Karel Neto\v{c}n\'{y}}
\email{netocny@fzu.cz}
\affiliation{Institute of Physics, Czech Academy of Sciences, Prague, Czech Republic}

\begin{abstract}

	We consider stationary driven systems in contact with a thermal equilibrium bath.  There is a constant (Joule) heat dissipated from the steady system to the environment as long as all parameters are unchanged.  As a natural generalization from equilibrium thermodynamics, the nonequilibrium heat capacity measures the excess in that dissipated heat when the temperature of the thermal bath is changed.   To improve experimental accessibility we show how the heat capacity can also be obtained from the response of the instantaneous heat flux to small periodic temperature variations.
	\end{abstract}

\maketitle

\section{Calorimetry out of equilibrium}

Standard thermodynamics deals with equilibrium systems and their energy exchange with the environment as external parameters like the volume or the temperature are changed. If these changes are slow (quasistatic and along equilibria), then the entire scheme simplifies as described by the thermodynamic laws for reversible processes.  There heat and entropy get proportional (Clausius heat theorem).  As the ability of a specific system to exchange heat and store energy is usually given in terms of the heat capacity (the heat contribution from temperature changes) and the latent heat (when temperature is unchanged), those quantities also yield important information about the internal structure of the equilibrium system.  In that way, calorimetry has provided crucial information about microscopic structures and the nature of the physical states.  
 
Concerning possible generalizations, the first (and by now more standard) option is to go beyond quasistatic processes and towards a time-resolved thermodynamics. That can be done within the framework of linear response and it naturally leads to a frequency-dependent generalization of the heat capacity and related thermodynamic quantities. That was mostly studied for equilibrium systems, with the zero-frequency limit recovering the usual reversible thermodynamic picture.  See e.g. \cite{ND} and also the very recent \cite{dio} in the context of stochastic thermodynamics.
In the present paper we suggest a similar step forward in the study of thermodynamic processes connecting steady states of nonequilibrium systems.\\ 

It is important to realize that in driven nonequilibrium systems the heat exchange runs upon a dissipation background. It means that the total heat exchanged with the environment goes virtually to infinity when making the process slower and slower because of its DC-component coming from the steady dissipation. Hence, we are really interested not in the (eventually diverging) total heat but in its \emph{excess} part coming from changes in the temperature and/or other parameters.

On the theoretical side, a natural question arises whether such an excess heat is well defined in the quasistatic regime, in the sense of being essentially insensitive of the actual speed of the process as long as it is slow enough. This question has been answered in the affirmative; see \cite{eu,jir}. It allows to consistently construct a generalization of the heat capacity to nonequilibrium steady states.   We have checked via examples that such a steady heat capacity exhibits some new features when far from thermal equilibrium. For example, it can take negative values. Nevertheless, some more systematic understanding of how these properties reflect the structure of nonequilibrium steady states is still lacking.  Response to temperature variations in the general context of fluctuation--dissipation relations has also been discussed in \cite{yol}.
   
Towards the experimental realization, there are other problems. First, one may want to measure the excess heat directly along a relaxation process to the new steady condition after making a small sudden change of temperature (or other parameters). One then needs to extract the transient part of the dissipated heat, i.e., the one obtained after subtracting the steady ``background'' dissipation. As a possible variation, instead of measuring the heat directly, it can also be accessed indirectly from measuring the (excess) work done by the driving forces. Yet, main issues to be solved here include the finding of the experimentally most feasible systems on which the temperature can be manipulated on time scales comparable with those of the system itself.  The present paper seeks an alternative route: to extend the frequency-dependent calorimetry to truly nonequilibrium systems and to extract the quasistatic excess from its low-frequency behavior.  That is the main purpose of the present paper.\\

We start in the next Section with the definition of (nonequilibrium) heat capacity.  We also include in Section \ref{mar} some relevant formulae how to rewrite that specifically for processes modeled as Markov dynamics, in terms of dissipated power.  The main result of the paper is in Section \ref{di} which describes the method of measuring heat capacity via temperature modulation and for which we believe the problem of excess (as a difference between very large quantities) may be avoided.  Instead of making the difference of time-extensive heats, we consider there the heat flux as function of time.  The heat capacity of nonequilibrium steady states then also appears as the static
limit of a nonequilibrium frequency-dependent heat capacity.

\section{Nonequilibrium theory}

We refer to \cite{eu,jir} for the initial theory and basic examples of nonequilibrium heat capacities. The basic idea builds on concepts from steady state thermodynamics as in~\cite{oon,kom2}.
The result of the present paper is to see in Section \ref{di} that  the specific heat of a system under  steady dissipative conditions can be measured by following the dissipated power as a function of time.
We start however next with the basic formul{\ae} which rigorously connect the nonequilibrium heat capacity with the excess heat. 

\subsection{Quasistatic excess heat}

Consider a generic thermodynamic system on which external forces perform some work $W$ and which exchanges heat $Q$ with an (equilibrium) heat bath at a temperature $T$, so that
$W + Q = \De U$ is the energy balance. We assume that the external forces maintain the system under fixed nonequilibrium conditions before time zero, so that they perform work
$W_{[-t,0]} = w^{(T)}\,t$ at constant power $w^{(T)} > 0$, which passes through the system and then dissipates as heat
$-Q_{[-t,0]} = -q^{(T)}\,t$ at rate $-q^{(T)} = w^{(T)}$. 
We explicitly indicate the dependence on the temperature $T$ playing the role of a control parameter. We remark that this always means the (well-defined) temperature of the equilibrium heat bath to which the system dissipates.

Both heat and work are time-extensive and nonzero because of assumed nonequilibrium conditions.  In applications to fluctuating mesoscopic systems, heat and work can be physically well-defined per trajectory when the system is weakly coupled to the environment, but the heat capacity involves taking statistical averages
over possible system trajectories; see Appendix~\ref{mar}.\\ 

Assume now that we make a measurement of the heat under slow temperature changes starting from time zero. A general quasistatic process can be decomposed in many elementary processes, each one consisting of a tiny sudden warming up (or cooling down) and then followed by a relaxation to new steady conditions. We could sum all the elementary contributions but clearly, for both theoretical and experimental purposes, it is enough to concentrate on one such an elementary process.

Before the sudden change of temperature from $T$ to $T + \de T$ at time zero, the system was in the steady state corresponding to the bath temperature $T$. After the change, it undertakes a relaxation to the new steady state at $T + \de T$. That is a transient process and the heat
$Q_{[0,t]}$  is no longer purely extensive but it contains a transient part as well. The latter can be extracted by comparing with the steady heat under the new stationary conditions, which is $q^{(T+\de T)}\,t$. That transient contribution along the complete relaxation process,
\begin{equation}\label{qex}
  \de Q^{\text{ex}} = \lim_{t \to \infty} \Bigl( Q_{[0,t]} - q^{(T+\de T)}\,t \Bigr)
\end{equation}
is called an \emph{excess heat}. Note that we really have to subtract the steady heat as corresponding to the \emph{new} temperature $T + \de T$ since the dissipation rate can be (and typically is) temperature-dependent. Under equilibrium conditions the latter would be just zero and the excess heat coincides with the total heat exchange along the elementary process. In contrast, out of equilibrium we take the difference of large (in the limit, infinite) quantities. In practice, one surely performs no time limit but, instead, let the relaxation run till it is ``essentially finished''. If $\tau$ is a characteristic time of relaxation then the excess heat $\de Q^\text{ex}$ is to be compared with the steady heat $q^{(T)} \tau$. Obviously, if 
$|q^{(T)}|\tau \gg |\de Q^\text{ex}|$ then one can hardly expect the excess heat to be distinguishable against the steady dissipation background. 

\subsection{Steady heat capacity}

The steady heat capacity quantifies the extra heat needed for the system to accommodate to a unit temperature change,
\begin{equation}\label{hca}
C(T) = \frac{\de Q^\text{ex}}{\de T}
\end{equation}
 Analogously, one can consider more general quasistatic processes including also the change of other thermodynamic parameters, which would then lead to a nonequilibrium generalization of the latent heat (capacities). All these quantities naturally supplement the incoming heat flux $q^{(T)}$ and provide a more complete characterization of the nonequilibrium steady state and its thermal sensitivity to external perturbations.\\

Although heat is a primary quantity here, we can as well consider the \emph{excess work} defined analogously as
\begin{equation}\label{wex}
  \de W^\text{ex} = \lim_{t \to \infty} \Bigl( W_{[0,t]} - w^{(T+\de T)}\,t \Bigr)
\end{equation}
where always $W_{[0,t]} + Q_{[0,t]} = U(t) - U(0)$.
Since the steady power on the system is just 
$w^{(T+\de T)} = -q^{(T+ \de T)}$, we can relate~\eqref{wex} with~\eqref{qex} in the balance
$\de W^\text{ex} + \de Q^\text{ex} = \id U$. Hence, the steady heat capacity \eqref{hca} can also be written in the form
\begin{equation}\label{tex}
  C(T) = \frac{\partial U}{\partial T} - \frac{\de W^\text{ex}}{\de T}
\end{equation}
where the first term is a usual temperature-energy response. Under equilibrium conditions such as constant volume and/or other thermodynamic coordinates, the second term vanishes. In this case the familiar equilibrium formula is recovered, namely that the equilibrium heat capacity coincides with the temperature-energy response coefficient. In contrast, the nonequilibrium contribution cannot be reduced to such a simple ``thermodynamic'' form and it depends on dynamical details of the system.\\

In Appendix~\ref{mar} we derive an explicit form of the nonequilibrium heat capacity for general Markov systems obeying the local detailed balance principle. The result reads that besides the steady-state average energy 
$U = \langle E(x) \rangle_T$, with $E(x)$ the energy function on mesoscopic states $x$, we need still another function 
$V^T(x)$, $\langle V^T(x) \rangle_T = 0$, which encapsulates the effect of nonequilibrium driving forces. In total,
\begin{equation}\label{c-mar}
C(T) = \frac{\id \langle E(x) \rangle_{T}}{\id T} - 
\Bigl\langle \frac{\id V^{T}(x)}{\id T}\, \Bigr\rangle_T
\end{equation}

An important feature of the new function $V^T(x)$ is that it depends both on the state $x$ and the bath temperature $T$. Suppose we can approximately write 
$V^T(x) \simeq \Phi(x) - \langle \Phi(x) \rangle_T$ with a temperature-independent ``potential'' $\Phi(x)$. Then
\begin{equation}
C(T) \simeq \frac{\id \langle \tilde E(x) \rangle_{T}}{\id T}\,,\qquad
\tilde E(x) = E(x) + \Phi(x)
\end{equation}
and we obtain an approximate formula resembling the equilibrium form for the modified energy function $\tilde E(x)$. Indeed, this is a viable simplification, e.g., in the regimes of very low or very high temperature, see~\cite{jir} for specific examples. However, such a decomposition of the function $V^T(x)$ is not possible in general.

\section{Temperature--heat response}\label{di}

In order to overcome possible experimental difficulties with measuring the excess heat above the steady dissipation background, we next discuss an alternative but theoretically equivalent scenario within the framework of time-resolved calorimetry. 

The heat can generally be resolved into the time-dependent flux as
$Q_{[0,t]} = \int_0^t J^Q_s\,\id s$. 
Initially we have the steady heat current $J_0^Q  = q^{(T)}$ into the system (equal to minus the steady rate of dissipation at temperature $T$).  Let us now modulate the temperature, $T_s = T + h_s$, at times $s \geq 0$. Within the linear response theory the heat current at time $t>0$ is
\begin{equation}\label{adm}
J^Q_t = J^Q_0 + \lambda_\infty\,h_t + \int_0^t \lambda_s\,h_{t-s}\,\id s
\end{equation}
The function $\lambda_t$ is a temporal temperature-heat ``admittance'', assumed to decay fast enough in time;  $\lambda_\infty$ accounts for the immediate, non-delayed response. The latter naturally emerges in Markov systems with discrete states as a consequence of temporal coarse-graining; for a more general discussion on delayed and non-delayed contributions in the linear theories see, e.g., Section 3.1.2 in \cite{kubo}. 
For other considerations of fluctuation-response relations for thermal perturbations in overdamped diffusions, see~\cite{yol}. \\

Let us again take the special case where the temperature suddenly changes at time zero from $T$ to $T+\de T$ (i.e., $h_s = \de T$ for $s>0$).  
In the limit $t \to \infty$ the system approaches a new steady state at bath temperature
$T + \de T$ with the steady heat current 
$J_t^Q \rightarrow J_\infty^Q = q^{(T+\de T)}$. 
From~\eqref{qex}--\eqref{hca} the heat capacity $C(T)$ satisfies
\[
\int_0^\infty ( J^Q_t - J^Q_\infty)\,\id t = C(T)\,\de T
\]
so that \eqref{adm} yields
\begin{equation}\label{c-ka-relation}
C(T) = -\int_0^\infty \id t \int_t^\infty \lambda_s\,\id s = -\int_0^\infty t \lambda_t\,\id t
\end{equation}
which expresses the heat capacity in terms of the admittances. On the other hand, the shift in steady heat currents is, again from \eqref{adm},
\begin{equation}\label{sig}
J_\infty^Q - J_0^Q = B(T) \,\de T\,,\qquad
B(T) = \lambda_\infty + \int_0^\infty \lambda_t\,\id t
\end{equation}
This way both response coefficients $B(T)$ and $C(T)$ derive from the admittance $\lambda_s$ and they capture different aspects of the temperature-heat response.

As a more experimentally feasible protocol we consider the harmonic temperature oscillations
$h_s = \epsilon \,\sin(\omega s)$ with some small amplitude $\epsilon$ and frequency $\omega$. Provided the admittance $\lambda_s$ decays asymptotically as $O(e^{-\gamma s})$ with some $\gamma > 0$, the heat current at large times obtains the form
\begin{equation}\label{gaga}
J^Q_t = q^{(T)} + \epsilon\,[\si_1(\om) \sin(\om t) + \si_2(\om) \cos(\om t)] + O(e^{-\ga t})
\end{equation}
defining $\si_{1,2}(\om)$ as the in- and out-phase components of the temperature-sensitivity of the dissipation. Comparing with \eqref{adm}, they are related to the admittance $\lambda_t$ by the Fourier-Laplace transform
\[
  \si_1(\om) + i\, \si_2(\om) = \lambda_\infty + \int_0^\infty e^{-i\om\,t}\lambda_t\,\id t
\]
From \eqref{sig} we get
$\si_1(\om = 0) = B(T)$,\,  $\si_2(\om = 0) = 0$,\, and from \eqref{c-ka-relation},
\begin{equation}
  \frac{\partial\si_1}{\partial\om}\Bigr|_{\om = 0} = 0\,,\qquad
  \frac{\partial\si_2}{\partial\om}\Bigr|_{\om = 0} = C(T)
\end{equation}
Hence, combining that with \eqref{gaga}, the low-frequency asymptotics of the heat current response is
\begin{equation}
  J^Q_t = J_0^Q + \epsilon\,[B(T)\, \sin(\om t) + C(T)\,\om \cos(\om t) + O(\om^2)] + O(e^{-\ga t})
\end{equation}
We see that the nonequilibrium heat capacity, as originally defined via the excess heat,  provides the leading low-frequency (out-phase) correction to the steady (in-phase) linear temperature-heat relation. This also indicates how the steady heat capacity can be detected and measured from the response to slow periodic temperature variations.

Note that this is nothing but a frequency-dependent calorimetry restricted to low frequencies, see e.g.~\cite{dio}, the only difference being that in the usual equilibrium setup $J_0^Q= q^{(T)}$ vanishes. In contrast, around a steady nonequilibrium the latter provides the dominant (for $\omega \to 0$) contribution to the heat flux, whereas the heat capacity becomes the next correction. 

\section{Conclusions}
 Thermal properties of nonequilibria appear essential in the program of steady state thermodynamics.
 Calorimetry of nonequilibrium systems may be developed to provide a useful characterization of the change in a material's thermal properties when driven away from equilibrium conditions, \cite{bioc,b2}. 
Nonequilibrium heat capacities can be consistently defined in terms of the notion of excess heat, or from how the steady dissipated power varies with temperature. We have seen how temperature modulation for nonequilibria gives access to that information via the time-dependence of the instantaneous heat flux.

\vspace{2mm}
\noindent {\bf Acknowledgments:}  
KN acknowledges the support from the Grant Agency of the Czech Republic, grant no.~17-06716S.


\appendix
\section{Heat capacity of Markov systems}\label{mar}

In this section we derive formula~\eqref{c-mar} for the nonequilibrium heat capacity of Markov systems with discrete states, which are often used as  models in stochastic thermodynamics. For further details see~\cite{eu,jir}.  

We consider a system with finitely many states $x$ which are uniquely associated to an energy level $E(x)$. The system is in contact with an equilibrium bath at temperature $T$ and it is also driven by external forces. It means that whenever there occurs a transition
$x\rightarrow y$, the driving forces perform some work 
$\caW(x,y) = -\caW(y,x)$ on the system and some heat 
$\caQ(x,y)$ enters the system from the bath. They are related by the energy balance
\begin{equation}
E(y) - E(x) = \caW(x,y) + \caQ(x,y)
\end{equation}
As we want to model a genuine nonequilibrium system, we assume that $\caW(x,y)$ cannot be written as a difference of some potential which could then be included in the energy function $E(x)$.

The Markov dynamics is introduced via transition rates $k^T(x,y)$ for each admissible transition 
$x \rightarrow y$. For thermodynamic consistency, they have to satisfy the local detailed balance principle~\cite{der,mn},
\begin{equation}
\frac{k^T(x,y)}{k^T(y,x)} = \exp \Bigl[ -\frac{\caQ(x,y)}{k_B T} \Bigr]
\end{equation}
In particular, the rates depend on the bath temperature as indicated in our notation. 

Recall that we want to compute the (average) excess work~\eqref{tex} along the relaxation process started from the steady state distribution at bath temperature $T$ and then running under the dynamics corresponding to the temperature $T + \de T$. The work 
 done by driving forces can be obtained for any trajectory of the system by summing up contributions $\caW(x_j,x_{j+1})$ from all subsequent transitions 
$x_j \rightarrow x_{j+1}$ along that trajectory. To find its statistical average it is useful to introduce the instantaneous 
power: Given the system at $x$ and attached to the equilibrium bath at temperature $T + \de T$, the average power of driving forces, i.e.~the work per unit time, is
\begin{equation}
\caP^{T+\de T}(x) = \sum_{y \neq x} k^{T+\de T}(x,y)\,\caW(x,y)
\end{equation}
Then the average work is
\begin{equation}
W_{[0,t]} = \Bigl\langle \int_0^t \caP^{T+\de T}(x_t)\,\id t \Bigr\rangle_{T \rightarrow T+\de T}
\end{equation}
where $\langle \cdot \rangle_{T \rightarrow T+\de T}$ stands for averaging with respect to the process started from the steady state at $T$ at $t=0$ and then running dynamics with the transition rules 
$k^{T+\de T}(x,y)$. 
Analogously, the steady state power is given by the stationary average
\begin{equation}
w^{(T+\de T)} = \langle \caP^{T+\de T}(x) \rangle_{T + \de T} =
\sum_x \caP^{T+\de T}(x)\,\rho_{T + \de T}(x)
\end{equation}
where $\rho_{T + \de T}(x)$ is the stationary distribution given the bath is at temperature $T + \de T$. Hence the excess work is
\begin{equation}
\begin{split}
\de W^\text{ex} &= \Bigl\langle \int_0^\infty 
\Bigl[ \caP^{T+\de T}(x_t) - \bigl\langle \caP^{T+\de T}(x) \bigr\rangle_{T + \de T} \Bigr]\,\id t \Bigr\rangle_{T \rightarrow T+\de T}
\end{split}
\end{equation}
This can be somewhat simplified by introducing the function
\begin{equation}
V^{T+\de T}(x) = \Bigl\langle \int_0^\infty 
\Bigl[ \caP^{T+\de T}(x_t) - \bigl\langle \caP^{T+\de T}(x) \bigr\rangle_{T + \de T} \Bigr]\,\id t \, \Bigl| \bigr. \, x_0 = x \Bigr\rangle_{T+\de T}
\end{equation}
where the conditional average means that the process starts from $x$ and then runs according to the dynamics at $T + \de T$. Note that it now depends only on a single temperature 
(in this case $T + \de T$), and by construction $\langle V^{T}(x) \rangle_{T} = 0$ for any $T$. This finally yields
\begin{equation}
\de W^\text{ex} = \langle V^{T+\de T}(x) \rangle_T = 
\Bigl\langle \frac{\id V^T(x)}{\id T} \Bigr\rangle_T \de T
\end{equation}
Together with $U = \langle E(x) \rangle_T$ we obtain 
\begin{equation}
C(T) = \frac{\id \langle E(x) \rangle_T}{\id T} - 
\Bigl\langle \frac{\id V^T(x)}{\id T} \Bigr\rangle_T 
\end{equation}
which is formula~\eqref{tex}. 

A more rigorous derivation employs the quasistatic limit of any smooth time dependence of temperature, see~\cite{eu}. The function $V^T(x) $ can be conveniently computed in terms of the backward Kolmogorov generator, see~\cite{jir}. 
For diffusion processes similar expressions hold, as for example made explicit in Eq.~(3.5) in~\cite{mcl}.  For dissipative mechanical systems we have the usual 
expression for the power ${\cal P}^T(p,q) = F(q)\cdot p$  for nonconservative force $F$ (which can depend implicitly on $T$) and states $x=(p,q)$ in  phase space.

Close to equilibrium, when $\caW(x,y) = \ve \caW_1(x,y)$ with $\ve$ a small parameter, more explicit expressions for the heat capacity $C(T)$ can be obtained by invoking McLennan ensembles to approximate $\langle\cdot\rangle_T-$expectations, see~\cite{mcl}.  As it happens, in linear order around equilibrium (up to order $\ve$), the correction to the Gibbs ensemble is exactly given by $V^T$:
\[
\rho_T(x) = \frac{1}{Z} \exp \{-\beta [E(x) + V^T(x) + O(\ve^2)]\}
\]
Per consequence, close to equilibrium 
the heat capacity is given by
\begin{equation}\label{thre}
C(T) = \frac{\id \langle E(x) \rangle_T}{\id T}\ -
\frac{\langle E(x) \rangle_T -\langle E(x) \rangle_T^{\text{eq}}}{T} + 
O(\ve^2)
\end{equation}
where $\langle \cdot \rangle^\text{eq}$ is the average under the equilibrium Gibbs ensemble $(\ve = 0)$.


\begin{thebibliography}{10}
	
	\bibitem{ND}
	J.~K.~Nielsen and J.~C.~Dyre,
	Fluctuation-dissipation theorem for frequency-dependent specific heat.
	\emph{Phys. Rev. B} \textbf{54}, 15754 (1996).
	
	\bibitem{dio}
	M.~J.~de Oliveira, Complex heat capacity and entropy production of temperature modulated systems. arXiv:1905.10306v1 [cond-mat.stat-mech]
	
	\bibitem{eu}
	E.~Boksenbojm, C.~Maes, K.~Neto\v{c}n\'{y}, and J.~Pe\v{s}ek, Heat capacity in nonequilibrium steady states.
	\emph{Europhys. Lett.} \text{96}, 40001 (2011).
	
	\bibitem{jir}
	J.~Pe\v{s}ek, E. Boksenbojm, and K.~Neto\v{c}n\'{y},
	Model study on steady heat capacity in driven stochastic systems.
	Cent. Eur. J. Phys. {\bf 10}(3), 692--701 (2012).
	
	\bibitem{yol}
	C.~Yolcu et al, A general fluctuation-response relation for noise variations and its application to driven hydrodynamic experiments. 
	\emph{J. Stat. Phys.} \textbf{167}, 29 (2017).
	
		\bibitem{oon}
	Y. Oono and M. Paniconi, Steady state thermodynamics. \emph{Prog. Theor. Phys. Suppl.} {\bf 130}, 29 (1998).
	
	\bibitem{kom2}
	T.~S.~Komatsu, N.~Nakagawa, S.-I.~Sasa and H.~Tasaki,
	Steady State Thermodynamics for Heat Conduction -- Microscopic Derivation.
	\emph{Phys. Rev. Lett.}, {\bf 100}, 230602 (2008).\\---,
	\emph{J. Stat. Phys.} {\bf 134},  401 (2009).
	
	\bibitem{kubo}
		R.~Kubo, M.~Toda, and N.~Hashitsume, \emph{Statistical Physics II: Nonequilibrium statistical mechanics} (Springer-Verlag, New York, 1986).
		
	
	\bibitem{bioc}
	{\it Handbook of Thermal Analysis and Calorimetry: From macromolecules to man},
	Volume 4, Eds.Patrick Kent Gallagher, Michael E. Brown and Richard B. Kemp, Elsevier, 1999.
	
	\bibitem{b2}
	{\it The nature of biological systems as revealed by thermal methods},
	Volume 5 of Hot topics in thermal analysis and calorimetry, Ed. D\'enes L\"orinczy, Springer, 2004.
	
	

	\bibitem{der}
	B.~Derrida, Non-equilibrium steady states: fluctuations and large deviations of the density and of the current, 
	\emph{J. Stat. Mech.}, P07023 (2007).
	
	\bibitem{mn}
	C. Maes and K. Neto\v{c}n\'y, Time-reversal and entropy
	\emph{J. Stat. Phys.} \textbf{110}, 269 (2003).
	
	\bibitem{mcl}
		C.~Maes and K.~Neto\v{c}n\'y,
		Rigorous meaning of McLennan ensembles.
		\emph{J. Math. Phys.} {\bf 51}, 015219 (2010).
		
\end{thebibliography}
\end{document}